\documentclass[sigconf]{acmart}

\acmBadge[https://doi.org/10.5281/zenodo.10513837]{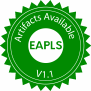}
\acmBadge[https://doi.org/10.5281/zenodo.10513837]{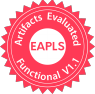}

\usepackage[T1]{fontenc}
\usepackage{graphicx}
\usepackage{amsmath}
\usepackage{amsfonts}
\usepackage{subcaption}
\usepackage{booktabs}
\usepackage{appendix}
\usepackage{tabularx}
\usepackage{tabularx}
\newcolumntype{Y}{>{\centering\arraybackslash}X}

\usepackage[noend]{algpseudocode}
\usepackage{algorithm}
\algblock{ParFor}{EndParFor}
\algnewcommand\algorithmicparfor{\textbf{parfor}}
\algnewcommand\algorithmicpardo{\textbf{do}}
\algnewcommand\algorithmicendparfor{\textbf{end\ parfor}}
\algrenewtext{ParFor}[1]{\algorithmicparfor\ #1\ \algorithmicpardo}
\algrenewtext{EndParFor}{\algorithmicendparfor}
\algnewcommand\algorithmicinput{\textbf{Input:}}
\algnewcommand\algorithmicoutput{\textbf{Output:}}
\algnewcommand\Input{\item[\algorithmicinput]}%
\algnewcommand\Output{\item[\algorithmicoutput]}%

\usepackage{color}
\usepackage{multirow}
\usepackage{url}

\usepackage{array}
\newcolumntype{P}[1]{>{\centering\arraybackslash}p{#1}}

\usepackage{xcolor}

\newcommand{\classes}[0]{Y}
\newcommand{\class}[0]{y}
\newcommand{\numClass}[0]{N}

\newcommand{\figref}[1]{Figure~\ref{#1}}
\newcommand{\tabref}[1]{Table~\ref{#1}}
\newcommand{\secref}[1]{Section~\ref{#1}}
\newcommand{\defref}[1]{Definition~\ref{#1}}

\begin{document}
\title{Case Study: Neural Network Malware Detection Verification for Feature and Image Datasets}

\author{Preston K. Robinette}
\email{preston.k.robinette@vanderbilt.edu}
\affiliation{%
  \institution{Vanderbilt University}
  \city{Nashville}
  \state{TN}
  \country{USA}
}

\author{Diego Manzanas Lopez}
\email{diego.manzanas.lopez@vanderbilt.edu}
\affiliation{%
  \institution{Vanderbilt University}
  \city{Nashville}
  \state{TN}
  \country{USA}
}

\author{Serena Serbinowska}
\email{serena.serbinowska@vanderbilt.edu}
\affiliation{%
  \institution{Vanderbilt University}
  \city{Nashville}
  \state{TN}
  \country{USA}
}

\author{Kevin Leach}
\email{kevin.leach@vanderbilt.edu}
\affiliation{%
  \institution{Vanderbilt University}
  \city{Nashville}
  \state{TN}
  \country{USA}
}

\author{Taylor T. Johnson}
\email{taylor.johnson@vanderbilt.edu}
\affiliation{%
  \institution{Vanderbilt University}
  \city{Nashville}
  \state{TN}
  \country{USA}
}
\copyrightyear{2024}
\acmYear{2024}
\setcopyright{acmlicensed}\acmConference[FormaliSE '24]{Formal Methods in
Software Engineering (FormaliSE)}{April 14--15, 2024}{Lisbon, Portugal}
\acmBooktitle{Formal Methods in Software Engineering (FormaliSE) (FormaliSE
'24), April 14--15, 2024, Lisbon, Portugal}
\acmDOI{10.1145/3644033.3644372}
\acmISBN{979-8-4007-0589-2/24/04}

\begin{abstract}

Malware, or software designed with harmful intent, is an ever-evolving threat that can have drastic effects on both individuals and institutions. Neural network malware classification systems are key tools for combating these threats but are vulnerable to adversarial machine learning attacks. These attacks perturb input data to cause misclassification, bypassing protective systems. Existing defenses often rely on enhancing the training process, thereby increasing the model's robustness to these perturbations, which is quantified using verification. While training improvements are necessary, we propose focusing on the verification process used to evaluate improvements to training. As such, we present a case study that evaluates a novel verification domain that will help to ensure tangible safeguards against adversaries and provide a more reliable means of evaluating the robustness and effectiveness of anti-malware systems. To do so, we describe malware classification and two types of common malware datasets (feature and image datasets), demonstrate the certified robustness accuracy of malware classifiers using the Neural Network Verification (NNV) and Neural Network Enumeration (nnenum) tools\footnote{Code, VNN-LIB: \url{https://github.com/pkrobinette/verify_malware}}, and outline the challenges and future considerations necessary for the improvement and refinement of the verification of malware classification. By evaluating this novel domain as a case study, we hope to increase its visibility, encourage further research and scrutiny, and ultimately enhance the resilience of digital systems against malicious attacks.

\end{abstract}

\maketitle

\section{Introduction}
Malware is software intentionally designed to cause damage, gain control, or disrupt an interface, network, or digital device. There are many use cases for malware, including stealing sensitive data (data theft), stealing data and holding it for ransom (ransomware attacks), the creation of a `botnet' (Distributed Denial of Service (DDoS)), using someone else's system to mine cryptocurrency (cryptojacking), monitoring activity and passwords (espionage), and distributing spam. Not only do malware attacks pose a significant threat to individuals, but they also frequently target a wide array of institutions. These include educational institutions like universities, businesses both private and public, healthcare entities, and governmental bodies. In a 2022 Response Report from Palo Alto Networks, reported ransomware payouts in the US reached up to \$8 million USD with demands received up to \$30 million USD as a result of ransomware attacks \cite{paloalto_report}. Malware represents a significant and ever-evolving threat to digital security, privacy, and integrity, and as such, is imperative to protect against.

\begin{figure}[t!]
    \centering
    \includegraphics[width=\columnwidth]{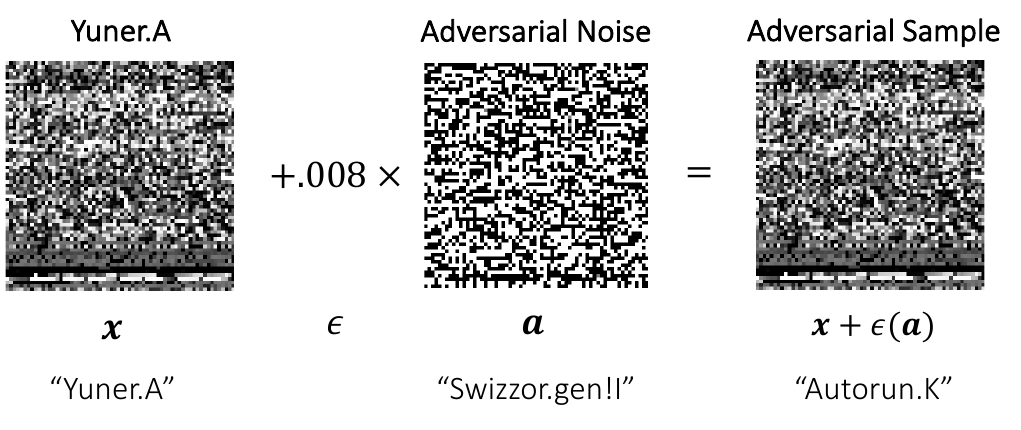}
    \caption{Adversarial machine learning example on Malimg dataset. Prior to the adversarial attack, the correct class ``Yuner.A'' is correctly predicted. This original malware image is then added with a small percentage $\epsilon$ of a noisy sample $a$ from a different class, resulting in the misclassification shown on the far right of the image ($x + \epsilon(a)$).}
    \label{fig:adv_ex}
\end{figure}

A key tool against this digital threat is the use of malware classification systems, which classify incoming and existing software. Trained models can detect malware in real-time, preventing systems from running the classified software, even if a user interacts with it. Most malware classification systems today utilize machine learning, and while adept at classification, they are also vulnerable to attacks. Bad actors can bypass the use of malware classification systems by utilizing adversarial machine learning. Adversarial machine learning relies on small perturbations in the input space to cause misclassification in the target system, as shown in \figref{fig:adv_ex}.  While adversarial attacks are applicable to all input types, this example demonstrates an adversarial attack on an image representation of a malware binary, which is a common classification technique in the malware domain.  In this example, prior to the adversarial attack, the correct class ``Yuner.A'' is correctly predicted. This original malware image is then added with a small percentage $\epsilon$ of a noisy sample $a$ from a different class, resulting in the misclassification shown on the far right of the image ($x + \epsilon(a)$).%
Adversarial attacks are not only imperceptible to the human eye, but they are also difficult to detect via anomaly detection because they are close to the original image. In \cite{carlini2017magnet}, the authors show that anomaly detection defenses such as variational autoencoders and generative adversarial networks are not robust to adversarial examples. Most defense mechanisms, therefore, rely on improvements in the training process, such as distillation \cite{papernot2016distillation}, adversarial retraining \cite{madry2018towards}, randomized smoothing \cite{carlini2023certified, lecuyer2019certified, cohen2019certified, yang2020randomized}, and randomized deletion \cite{huang2023rsdel}. Each of these methods increases the robustness of a trained model to perturbations in the input space, which can be evaluated quantitatively as the percentage of test inputs to be correctly classified within a bounded distance of a starting sample \cite{chen2020training}.

While the robustness of trained models has been used to verify classification networks in previous works, the focus has been on implementation improvements rather than on the domain itself. For instance, using robustness as a metric to show the improvements of using a robust training method as demonstrated on an image labeling dataset like CIFAR-10 \cite{cianfarani2022understanding}. Just as implementations improve under scrutiny, so too does a domain benefit from rigorous analysis, increased visibility, and continuous refinement, which is especially important for a domain as safety-critical as malware classification. 

Toward this end, we present an initial case study on malware classification verification and put forward the concept of `meaning' to a perturbation. For instance,  we acknowledge the possibility of applying an $\mathcal{L}_{\infty}$ perturbation on a binary feature; however, this does not yield any significant or meaningful semantic implications to the original binary file. These efforts work towards providing tangible guarantees against bad actors seeking to bypass classification schemes and provide valuable insight into a model's reliability to protect safety-critical networks. To do so, we describe two common malware dataset types and demonstrate the certified robustness accuracy of two types of malware classifications using the Neural Network Verification (NNV) ~\cite{tran2020cavtool,manzanas2023cav} and Neural Network Enumeration (nnenum) \cite{bak2021nfm} tools. Through this process, we identify domain challenges and areas to be considered moving forward. The contributions of this work are the following: %
\begin{enumerate}
    \item Novel application of formal verification to neural network malware classification (binary and family). 
    \item Identify `meaningful' perturbations for two different types of malware datasets (feature and image) and classification types (binary and family).
    \item Outline the challenges and future considerations necessary for the improvement and refinement of the verification of neural network malware classification.
\end{enumerate}%
\section{Preliminaries}
In this section, we introduce the malware datasets used in this work, the types of malware classification to which the datasets are applied, NNV and nnenum tools for certifying the robustness of trained models, and the metrics used to evaluate trained models.%
\subsection{Malware Datasets}
Below, we discuss two commonly-used types of malware datasets: 
(1) feature datasets, which represent malware samples as vectors of data such as byte entropy and string length, and 
(2) image datasets, which represent malware samples as grayscale images.

\subsubsection{Feature Datasets}
\label{sec:bodmas}
Malware feature datasets are typically composed of characteristics or \emph{features} extracted from a variety of malware samples. The features can include static attributes such as file size, header information, and/or the presence of specific strings, as well as dynamic attributes that are revealed when the malware is executed in a controlled environment, such as API calls, network activity, and/or changes to the file system or registry. %
\begin{table*}[tbh!]
    \centering
    \caption{BODMAS dataset feature types and examples. The ranges of values for each feature type are drastically different, showing the importance of using range-specific perturbations during verification.}
    \label{tab:bodmas_ftype}
    \begin{tabular}{P{0.2\textwidth}P{0.1\textwidth}P{0.18\textwidth}P{0.18\textwidth}p{0.28\textwidth}}
        \toprule
         \textbf{Feature Type} & \textbf{Count} & \textbf{Max Range Pre-Scale} & \textbf{Max Range Post-Scale} & \multicolumn{1}{c}{\textbf{Example}}\\
         \midrule
         Continuous & 5 & [5.0, 2.0e5] & [-0.1, 304.6] & Entropy of the file \\
         \hline
         Categorical & 8 & [0.0, 6.5e4] & [-0.0, 124.3] & Machine type \\
         \hline
         Hash Categorical & 560 & [-647.6, 15.4] & [0.0, 361.0] & Hash of original file\\
         \hline
         Discrete with Large Range & 34 & [0.0, 4.3e9] & [-0.0, 261.6] & Number of occurrences of each byte value within the file \\
         \hline
         Binary & 5 & [0.0, 1.0] & [-2.1, 0.5] & Presence of debug section\\
         \hline
         Hash Categorical Discrete & 1531 & [-8.0e6, 1.6e9] & [-327.9, 164.0] & Hash of target system type \\
         \hline
         Memory Related & 16 & [0.0, 4.0e9] & [-0.1, 307.5] & Size of the original file \\
         \hline
         Null & 222 & [-31.0, 60.0] & [-0.9, 160.4] & --- \\
         \bottomrule
    \end{tabular}
\end{table*}%
The extraction of these features can be conducted through a combination of static and dynamic analysis. 
Static analysis involves examining the malware code without executing it, and dynamic analysis involves running the malware in a sandboxed environment and observing and recording its behavior.

\textbf{BODMAS} The feature dataset used in this work is the Blue Hexagon Open Dataset for Malware Analysis (BODMAS), which is a collection of timestamped malware and benign samples for research purposes, co-created with Blue Hexagon \cite{bodmas}. It includes 57,293 malware (label 0) and 77,142 benign samples (label 1) collected between August 2019 and September 2020. The LIEF project was used to extract 2381 feature vectors from each sample using dynamic and static analysis, along with its classification label as benign or malicious. The extracted features can be broken down into seven different types, as shown in \tabref{tab:bodmas_ftype}. While the data can also be described by feature groups, we focus on data types as they influence the chosen specifications used for verification.

While features extracted with dynamic and static analysis offer valuable insights into a sample, they each have their own set of challenges. Static analysis can struggle against obfuscation techniques such as code encryption, packing, and the use of non-standard code constructs, which are used by malware authors to hide their code's true intent.  Additionally, static analysis can be time-consuming and resource-intensive when dealing with large and complex pieces of software and result in high false positive rates. Some sophisticated malware can even detect when they are being executed in a sandbox or virtual environment, altering their behavior or refusing to run altogether, thereby evading detection. 

\subsubsection{Image Datasets}
\label{sec:malimg}
\begin{figure}[tbh!]
    \centering
    \includegraphics[width=\columnwidth]{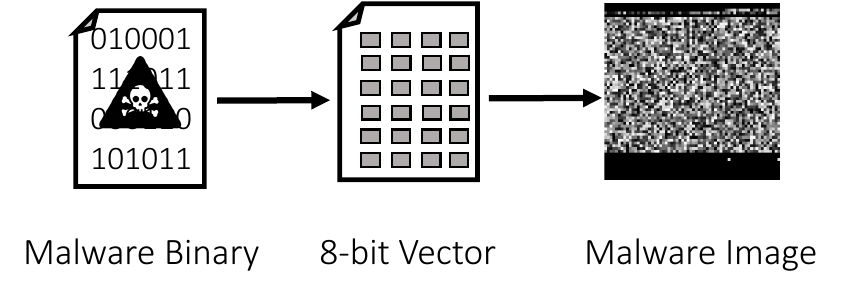}
    \caption{Malware binary conversion to an image. The malware binary is preprocessed to extract sequences of 8-bit vectors, or bytes. Each byte is then mapped to a pixel in the image, which is then plotted on a grid to create an image.}
    \label{fig:bin-to-im}
\end{figure}

\begin{figure*}[bth!]
    \centering
    \includegraphics[width=0.75\textwidth]{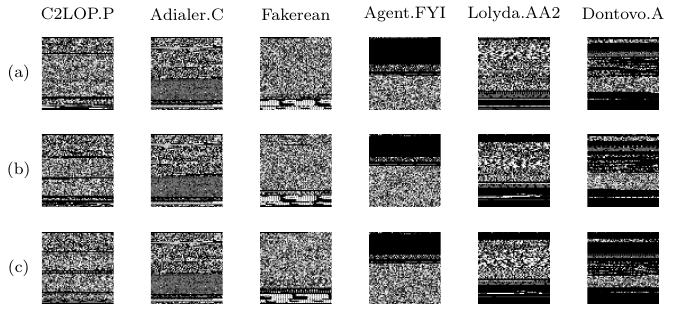}
    \caption{Malware family image examples. Each of the malware families produce relatively distinct patterns, making image classification an effective tool for malware detection and categorization. (a), (b), and (c) correspond to different samples of the same family indicated by the column name.}
    \label{fig:malimg_family}
\end{figure*}

Due to the difficulties of creating and maintaining up-to-date malware datasets using feature extraction, researchers also use image representations of software for training data \cite{singh2019malware, vasan2020image, kim2018image, kornish2018malware, nataraj2011malware}. This method works by converting binary executable files of malware into grayscale images, with each pixel in the image corresponding to a byte in the binary file, as shown in \figref{fig:bin-to-im}. These images can then be analyzed using image processing techniques or deep learning models originally designed for image recognition tasks, such as convolutional neural networks (CNNs). %
The process begins with the raw bytes of the malware binary being aggregated into an 8-bit vector. Each of these 8-bit vectors is then converted into a decimal value to fit the grayscale range (0-255). These values are then reshaped into a two-dimensional array, effectively creating an image. The resulting image, often referred to as a \emph{byteplot}, can reveal patterns that might be indicative of certain types of malware, as shown in \figref{fig:malimg_family}.%

\textbf{Malimg} The malware image dataset used in this work is the Malimg Dataset, which is composed of 9339 malware images from 25 different malware families \cite{nataraj2011malware}. The malware families are shown in \tabref{tab:malimg_per_class}.

\subsection{Malware Classification Types}
A malware classifier is an automated way to categorize and differentiate software (benign vs. malicious). In addition to the canonical classification of malicious vs. benign, malware classifiers are also used to identify a known malware family. A malware family refers to a group of malware that share significant characteristics and are often developed from the same code base. These shared characteristics can include similar behavior patterns, payload delivery methods, or purpose.  By identifying a sample's malware family, cybersecurity professionals gain insight into the nature, origin, purpose of the sample, threat level and type, propagation mechanisms, and potential damage. Additionally, successful identification can often lead to the identification of the creator's signature, which can help to identify the attack group responsible. All of these pieces of information aid in the subsequent development of effective countermeasures to neutralize or mitigate the effects of the detected malware. In this work, we train both \textbf{binary} classifiers ($Y_{\mathrm{binary}}=\{\mathrm{malicious}, \mathrm{benign}\}$) and \textbf{family} classifiers ($Y_{\mathrm{family}}=\{\mathrm{Adialer.C}, \mathrm{Agent.FYI}, \mathrm{Allaple.A}, ..., \mathrm{Yuner.A}\}$), where $Y$ is the set of all output classes.

\subsection{Neural Network Verification}
To analyze the neural network malware classifiers, we make use of NNV and nnenum tools. NNV is a software tool written in MATLAB\footnote{MATLAB [2022b]: \url{https://www.mathworks.com/products/new_products/r2022b-transition.html}} that implements reachability methods to formally verify specifications of neural networks (NN)~\cite{tran2020cavtool,manzanas2023cav}. NNV uses a star-set state-space representation and reachability algorithms that allow for a layer-by-layer computation of exact or overapproximate reachable sets for feed-forward (FFNN), convolutional (CNN)~\cite{tran2020cav}, Semantic Segmentation (SSNN)~\cite{tran2021cav}, and recurrent (RNN) \cite{tran2023hscc} neural networks, as well as neural ordinary differential equations \cite{manzanas2022formats}, and neural network control systems (NNCS) \cite{manzanas2022archcomp}. 

nnenum is a software tool written in Python that addresses the verification of ReLU neural networks through optimized abstraction refinement \cite{bak2021nfm}. This approach combines zonotopes with star set overapproximations for feed-forward (FFNN), convolutional (CNN), and recurrent (RNN) neural networks.
In this work, we focus on one type of property to verify: robustness, which is described in Definition~\ref{def:robustness}.%
\begin{definition}[Robustness]\label{def:robustness}
    Given a neural network classifier $f$, input $x \in \mathbb{R}^{n\times m}$, target $y \in \mathbb{R}^{\numClass}$ where $\numClass$ is the number of classes (i.e., $N=2$ for binary classification, $N=25$ for family classification), perturbation parameter $\epsilon \in \mathbb{R}$, and an input set $R$ containing $x_p$ such that $X_p = \{x: ||x - x_p|| \leq \epsilon\}$ which represents the set of all possible perturbations of $x$ where $||x - x_p||$ is the $\mathcal{L}_{\infty}$ norm. The classifier is locally \textbf{robust} at $x$ if it classifies all the perturbed inputs $x_p$ to the same label as $y$, i.e., the system is \textbf{robust} if $f(x_p) = f(x) = y$ for all $x_p \in X_p$.
\end{definition}

\subsection{Metrics}
\label{sec:metrics}
In this section, we introduce the classifier and verification performance metrics used in this work. 

\paragraph{\textbf{Classifier Performance}} To evaluate trained models prior to verification, we utilize accuracy, precision, recall, and F1 score metrics. We formalize these metrics across classes $\classes$ using \textit{True Positives} ($\text{TP}_{\class_i}$): samples \textbf{correctly} classified as class $\class_i$, \textit{True Negatives} ($\text{TN}_{\class_i}$): samples \textbf{correctly} classified as \textbf{not} in class $\class_i$, \textit{False Positives} ($\text{FP}_{\class_i}$): samples \textbf{incorrectly} classified as class $\class_i$, and \textit{False Negatives} ($\text{FN}_{\class_i}$): samples \textbf{incorrectly} classified as \textbf{not} in class $\class_i$, where $\class_i \in \classes$ s.t. $\classes = \{\class_1, \class_2, ..., \class_{\numClass}\}$ and $\numClass$ is the number of all classes in the dataset. The equations for accuracy, precision, recall, and F1 score are shown in Equations \ref{eq:acc}, \ref{eq:prec}, \ref{eq:rec}, and \ref{eq:f1} respectively.

\begin{equation}
\label{eq:acc}
\text{Accuracy} = \frac{\sum_{i=1}^\classes \text{TP}_{\class_i} + \text{TN}_{\class_i}}{\text{Total Samples}}
\end{equation}%
\begin{equation}
\label{eq:prec}
\text{Precision} = \frac{1}{N} \sum_{i=1}^\classes \frac{\text{TP}_{\class_i}}{\text{TP}_{\class_i} + \text{FP}_{\class_i}}
\end{equation}%
\begin{equation}
\label{eq:rec}
\text{Recall} = \frac{1}{N} \sum_{i=1}^\classes \frac{\text{TP}_{\class_i}}{\text{TP}_{\class_i} + \text{FN}_{\class_i}}
\end{equation}%
\begin{equation}
\label{eq:f1}
\text{F1} = 2 \cdot \frac{\text{Precision} \cdot \text{Recall}}{\text{Precision} + \text{Recall}}
\end{equation}%
\begin{algorithm}[htb!]
	\caption{Verification strategy}\label{alg:verify}
	\begin{algorithmic}[1]
        \small
		\Input{$f, x, target, \mathcal{A}, N_R$}  ~{\scriptsize \textcolor{blue}{ $\triangleright$ NN, input, label target, attack, \# of random examples}}
		\Output{res} ~{\scriptsize \textcolor{blue}{$\triangleright$ verification result}}
		\Procedure{\texttt{$res$ = malware\_verification}}{$f, x, \mathcal{A}, N_R$}
                \State $X_{rand}$ = $genRandExamples(x, \mathcal{A}, N_R)$ ~{\scriptsize \textcolor{blue}{$\triangleright$ generate random examples based on attack}}
                \For{{\texttt{$i=1:N$}}}
                    \State $y = f(X_{rand}(i))$ ~{\scriptsize \textcolor{blue}{$\triangleright$ classify random image}}
                    \If{$ y~\neq~target $}
                    \State $res = 0$ ~{\scriptsize \textcolor{blue}{$\triangleright$ counterexample found}}
                        \State \textbf{return} $res$ ~{\scriptsize \textcolor{blue}{$\triangleright$ stop procedure, no reachability analysis needed}}
                    \EndIf
                \EndFor
			\State $I = constructInputSet(x, \mathcal{A})$ ~{\scriptsize \textcolor{blue}{$\triangleright$ construct input set}}
			\State $R = Reach(f, I, relax)$ ~{\scriptsize \textcolor{blue}{$\triangleright$ compute reachable set using relax-star-area reachability}}
                \State $res = verify(R, target)$ ~{\scriptsize \textcolor{blue}{$\triangleright$ verify output set against target class}}
                \If{$res~\neq~1$} ~{\scriptsize \textcolor{blue}{$\triangleright$ if result is unknown, try again with approx}}
                    \State $R = Reach(f, I, approx)$ ~{\scriptsize \textcolor{blue}{$\triangleright$ compute reachable set using approx-star reachability}}
                    \State $res = verify(R, target)$ ~{\scriptsize \textcolor{blue}{$\triangleright$ verify output set against target class}}
                \EndIf
            \State \textbf{return} res  ~{\scriptsize \textcolor{blue}{$\triangleright$ return verification result $\{$0~=~ not robust, 1~=~robust, 2~=~unknown$\}$}}
		\EndProcedure
    \end{algorithmic}
\end{algorithm}%
\paragraph{\textbf{Verification Performance}}
The verification strategy implemented in NNV to certify the robustness of the BODMAS and Malimg benchmarks is described in Algorithm~\ref{alg:verify}. This algorithm consists of two main steps and three potential outcomes: %
\begin{enumerate}
    \item \textbf{Falsification} (0): In the first step of the verification process, we classify 500 random examples from the created input set (based on bound from adversarial attack). If a sample is misclassified, or a counterexample is found, this is considered a successful falsification, and the model is not robust to the perturbation $\epsilon$.
    \item \textbf{Robust} (1): If no counterexamples are found, then we proceed to run the reachability analysis: starting with an over-approximation method referred to as \emph{relax-approx-area} with a relax factor of 0.5. If the input set with \emph{relax-approx-area} evaluates to the correct class, the model is considered robust, and the valuation returns a 1.
    \item \textbf{Unknown} (2): If the verification result from \emph{relax-approx-area} is unknown, we proceed to run the \emph{approx-star} method, which is less conservative than the relax-star, but also more computationally expensive. If the sample with \emph{approx-star} is found to be robust, the valuation is 1; otherwise, the valuation is 2, which is considered unknown, as further refinement might or might not prove the model to be robust to perturbation $\epsilon$. 
\end{enumerate}%
In addition to this verification step, we also produce VNN-LIB files \cite{vnnlib23} for each dataset, sample, and epsilon value to be used by the nnenum verification tool. VNN-LIB is a standard file type to represent the specification of neural networks based on input-output relationships, derived from the Satisfiability Modulo Theories Library (STM-LIB)\footnote{SMT-LIB: \url{https://smtlib.cs.uiowa.edu}}. A VNN-LIB file consists of input bounds and a specification to verify. For instance, a VNN-LIB file for a Malimg dataset sample (64x64 image) with $\epsilon=2/255$ would consist of 4096 input variables (the pixels of the image) with a range $\pm 2/255$ of the original pixel value. The specification would then be that the true label is maintained (robust) for this range of inputs. As we need to create a different VNN-LIB file for each sample and epsilon value, there are 1200 VNN-LIB files for the BODMAS dataset (100 samples * 4 feature types * 3 epsilon values) and 375 VNN-LIB files for the Malimg dataset (125 samples * 3 epsilon values). A single VNN-LIB file and a corresponding trained model in an Open Neural Network Exchange (ONNX)\footnote{ONNX: \url{https://onnx.ai/}} format are then used as inputs to the nnenum tool, which presents three potential outcomes: Robust (\textit{holds}), Falsification (\textit{violated}), and Unknown (\textit{timeout}).

During verification, we evaluate models using two metrics: (1) average time $(s)$ to verify each input in the verification set for a given perturbation and (2) the number of inputs from the verification set for a perturbation ($\epsilon$) that are certified robust according to \defref{def:robustness} over the number of total samples, i.e., certified robustness accuracy (CRA). For each model that is tested, the verification set consists of randomly selected samples from the dataset. For instance, models from the BODMAS dataset are evaluated on 100 inputs. These 100 inputs are then used for verification in subsequent rounds, where each round corresponds to a particular perturbation size $\epsilon$. The average time metric would then be the average wall time it takes to verify across all 100 samples, and the CRA metric would be the number of these 100 samples that evaluate to Robust (1 for NNV, \textit{holds} for nnenum) over the number of total samples.

\section{Verification of Malware Classifiers Overview}
We aim to verify two different types of malware classifiers: binary classification (benign vs. malicious) and family classification (malware family). Each experiment contains three different phases: (1) training, (2) testing, and (3) verification.%
\begin{table}[tb!]
    \centering
    \caption{Neural network classifier model names and architectures used to train binary and family classifiers.
    }
    \label{tab:expr_params}
    \begin{tabularx}{\columnwidth}{XXp{3.5cm}}
        \toprule
         \multirow{2}{*}{\textbf{Classifier Type}} & \multirow{2}{*}{\textbf{Model Name}} & \textbf{Model Architecture} \\
         & & (in $\to$ [hidden layers] $\to$ out) \\
         \midrule
              \multirow{3}{=}{BODMAS\\(Binary)}          & none-2 & 2381 $\to$ 2  \\
               &4-2 & 2381 $\to$ [4] $\to$ 2  \\
               & 16-2 & 2381 $\to$ [16] $\to$ 2  \\
        \midrule
          \multirow{3}{=}{Malimg\\(Family)}                       & linear-25 & 4096 $\to$ 25 \\
          & 4-25 & 4096 $\to$ [4] $\to$ 25\\
          & 16-25 & 4096 $\to$ [16] $\to$ 25 \\
          \bottomrule
    \end{tabularx}
\end{table}%
\begin{table}[tb!]
    \centering
    \caption{Perturbations ($\epsilon$) used during verification.}
    \begin{tabular}{P{0.15\columnwidth} P{0.30\columnwidth} ccc}
    \toprule
         \textbf{Classifier Type} & \textbf{Coverage}  & \multicolumn{3}{c}{$\epsilon$} \\
         \midrule
          \multirow{4}{*}{BODMAS}& All & $0.01\%\phantom{00}$ & $0.05\%\phantom{00}$ & $0.1\%\phantom{00}$ \\
          & Discrete $\&$ Continuous &  $0.01\%\phantom{00}$ & $0.05\%\phantom{00}$ & $0.1\%\phantom{00}$ \\
          & Discrete &  $0.1\%\phantom{000}$ & $0.5\%\phantom{000}$ & $1\%\phantom{.000}$ \\
          & Continuous & $1\%\phantom{.0000}$ & $5\%\phantom{.0000}$ & $10\%\phantom{.000}$ \\
          \midrule
          Malimg & All & $\frac{1}{255}$ & $\frac{2}{255}$ & $\frac{3}{255}$\\
          \bottomrule
    \end{tabular}
    \label{tab:eps}
\end{table}%

\begin{table*}[tbh!]
    \centering
    \caption{Training results for BODMAS (binary) models and Malimg (family) models. Each of the models performs well according to accuracy, precision, recall, and F1 score metrics.}
    \label{tab:train_results_table}
    \begin{tabularx}{\textwidth}{XXXXXX}
        \toprule
         \textbf{Dataset} & \textbf{Model} & \textbf{Accuracy} & \textbf{Precision} & \textbf{Recall} & \textbf{F1} \\
         \midrule
        \multirow{3}{*}{BODMAS} & none-2 & 0.99 & 0.98 & 0.99 & 0.99 \\
                                & 4-2 & 0.99 & 0.99 & 0.99 & 0.99\\
                                & 16-2 & 0.99 & 0.99 & 0.99 & 0.99\\
        \midrule
        \multirow{3}{*}{Malimg} & linear-2 & 0.99 & 0.98 & 0.97 & 0.97\\
                                & 4-2 & 0.98 & 0.97 & 0.96 & 0.97\\
                                & 16-2 & 0.99 & 0.97 & 0.96 & 0.97\\
        \bottomrule
    \end{tabularx}
\end{table*}%
\begin{table*}[tbh!]
    \centering
    \caption{Certified robustness accuracy (CRA) and average time to verify verification results on 100 samples from the BODMAS (binary) dataset using NNV and nnenum verification tools on various feature types and ($\mathcal{L}_\infty$) perturbations.}
    \label{tab:bodmas_results}
\begin{tabularx}{\textwidth}{YXXYYY|YYY|YYY|YYY} 
\toprule 
\multirow{2}{*}{Metric} & \multirow{2}{*}{Model} & \multirow{2}{*}{Tool} &\multicolumn{3}{c}{All} & \multicolumn{3}{c}{Continuous \& Discrete} & \multicolumn{3}{c}{Discrete} & \multicolumn{3}{c}{Continuous} \\
 & & & 0.01 \% & 0.05 \% & 0.1 \% & 0.01 \% & 0.05 \% & 0.1 \% & 0.1 \% & 0.5 \% & 1 \% & 1 \% & 5 \% & 10 \% \\
 \midrule 
\multirow{6}{=}{\rotatebox{90}{CRA (\%)}} & \multirow{2}{*}{none-2} & NNV & 94 & 61 & 19 & 98 & 98 & 96 & 96 & 90 & 68 & 95 & 84 & 56 \\
 &  & nnenum & 94 & 61 & 19 & 98 & 98 & 96 & 96 & 90 & 68 & 95 & 84 & 56 \\
\cline{3-15} 
 \rule{0pt}{3ex} 
 & \multirow{2}{*}{4-2} & NNV & 99 & 69 & 26 & 100 & 100 & 99 & 100 & 98 & 88 & 100 & 97 & 81 \\
 &  & nnenum & 99 & 73 & 31 & 100 & 100 & 99 & 100 & 98 & 90 & 100 & 97 & 81 \\
\cline{3-15} 
 \rule{0pt}{3ex} 
 & \multirow{2}{*}{16-2} & NNV & 100 & 65 & 27 & 100 & 100 & 100 & 100 & 99 & 91 & 100 & 96 & 82 \\
 &  & nnenum & 100 & 71 & 34 & 100 & 100 & 100 & 100 & 99 & 92 & 100 & 96 & 82 \\
 \midrule
 \multirow{6}{=}{\rotatebox{90}{Avg. Time (s)}} & \multirow{2}{*}{none-2} & NNV & 0.35 & 0.30 & 0.29 & 0.17 & 0.17 & 0.17 & 0.27 & 0.27 & 0.28 & 0.31 & 0.26 & 0.19 \\
        &  & nnenum & 0.50 & 0.50 & 0.50 & 0.50 & 0.50 & 0.50 & 0.50 & 0.50 & 0.50 & 0.50 & 0.50 & 0.50 \\
        \cline{3-15} 
        \rule{0pt}{3ex} 
        & \multirow{2}{*}{4-2} & NNV & 0.31 & 0.42 & 0.62 & 0.17 & 0.17 & 0.17 & 0.28 & 0.28 & 0.31 & 0.29 & 0.27 & 0.28 \\
        & & nnenum & 0.50 & 0.50 & 0.51 & 0.50 & 0.50 & 0.51 & 0.50 & 0.50 & 0.51 & 0.50 & 0.50 & 0.51 \\
        \cline{3-15}
        \rule{0pt}{3ex}
        & \multirow{2}{*}{16-2} & NNV & 0.28 & 0.64 & 1.04 & 0.18 & 0.18 & 0.18 & 0.26 & 0.28 & 0.33 & 0.28 & 0.31 & 0.27 \\
        & & nnenum & 0.50 & 0.50 & 0.52 & 0.50 & 0.50 & 0.52 & 0.50 & 0.50 & 0.52 & 0.50 & 0.50 & 0.52 \\
\bottomrule
    \end{tabularx}
\end{table*}%
\subsection{Binary Classification}
\label{sec:bin_class}
To test the verification of binary malware classifiers (benign vs. malicious), we use the BODMAS dataset described in \secref{sec:bodmas}. Prior to training, the data is scaled using $z = \frac{x - \mu}{\sigma}$, where $z$ is the standardized score, $x$ is a given sample in a feature column, $\mu$ is the mean of the feature, and $\sigma$ is the standard deviation of the feature. The standard scalar standardization is used due to the large differences in value range between potential features.

Each of the remaining experimental phases is described in more detail below:%
\begin{enumerate}
    \item \textbf{Training:} Three different binary classifiers with varying architectures are trained using this data. The architectures used in these experiments are shown in 
    \tabref{tab:expr_params}. If a hidden layer is present, a ReLU activation function is used on the outputs of the hidden layer. These architectures are trained with sparse categorical cross-entropy loss function, Adam optimizer, 40 epochs, and a batch size of 128 samples. 
    \item \textbf{Testing:} To create a baseline for verification performance, we then test each classifier on the test data. From this, we gather accuracy, F1, precision, and recall metrics described in \secref{sec:metrics}. 
    \item \textbf{Verification:} We then use NNV and nnenum with the epsilon values shown in \tabref{tab:eps} to obtain the number verified and time to verify metrics for 100 randomly selected test samples. The randomly selected test samples are the same across all verification experiments. The randomly selected samples reflect the distribution of the data, which consists of $~43\%$ malicious samples. Additionally, as the BODMAS dataset consists of different types of features and ranges, the perturbation applied to each feature is the epsilon value percentage of the range for that particular feature. For instance, if feature 1 has a range $[3, 567]$, the $\mathcal{L}_{\infty}$ bound with $\epsilon=0.1\%$ would be $\pm0.56=0.1\%\times(567-3)$. This provides a more feature realistic perturbation to each sample.%
    \begin{enumerate}
        \item \textbf{All}: As a baseline, the epsilon perturbation (a) is applied to all 2381 features in a sample. While applying the epsilon perturbation to all features gives insight into the complete robustness of a model, a perturbation is not meaningful or impactful across all features. For instance, applying an $\epsilon = 0.1\%$ perturbation to a binary feature. While not all perturbed samples will be semantically meaningful, the $\mathcal{L}_{\infty}$ bound will contain some samples that are, providing a high-level evaluation of the model. 
        
        \item \textbf{Discrete \& Continuous}: From the available feature types, discrete and continuous features are the most applicable to $\mathcal{L}_{\infty}$ perturbations. As such, we apply epsilon perturbations (b) to only continuous and discrete features.
        
        \item \textbf{Discrete}: Discrete features offer insight into details like the number of imported libraries, the number of exported libraries, and the number of symbols in the file. As most of the discrete features involve counts of elements in the sample, small value changes can have a drastic impact on the classification of a sample. The epsilon values (c) chosen reflect this property.
        
        \item \textbf{Continuous}: In contrast to the discrete features, the continuous features generally result from calculations based on the elements in the sample, such as the entropy of strings or the average length of strings. These elements are more robust to perturbations, and larger epsilon values (d) were selected to test these elements.
    \end{enumerate}
\end{enumerate}%

\subsection{Family Classification}
To test the verification of malware family classifiers, we utilize the Malimg dataset described in \secref{sec:malimg}, which consists of image representations of malware samples from 25 different families. Prior to training, each of the images is resized to size 64x64, and the pixel values are normalized between 0 and 1. The data is then split into two different sets: training (8404 images) and testing (935 images).%
\begin{enumerate}
    \item \textbf{Training:} Three different family classifiers with varying architectures are trained using this data. The architectures used in these experiments are shown in the bottom part of \tabref{tab:expr_params}. If hidden layers are present, they represent a convolutional layer. Additionally, these convolutional models utilize a ReLU activation function. These architectures are trained with a categorical cross-entropy loss function, Adam optimizer, 10 epochs, and a batch size of 64 samples. 
    \item \textbf{Testing:} To create a baseline for verification performance, we then test each classifier on the test data. From this, we gather accuracy, F1, precision, and recall metrics.
    \item \textbf{Verification:} We then use NNV and nnenum with the epsilon values shown in \tabref{tab:eps} to obtain the number verified and time to verify metrics for 125 samples. To create this set, 5 samples from each class in the validation set are randomly selected. Each verification experiment utilizes the same 125 images. Out of the 25 classes in the validation data, 20 of them contain 25 images, giving the verification set $~20\%$ coverage on most classes. As the Malimg dataset consists of images, these epsilon values are applied to pixels in the sample image.
\end{enumerate}%
\section{Results}
In this section, we present the results of our evaluation. These experiments were conducted on a macOS with a 2.3
GHz 8-Core Intel Core i9 processor with 16 GB 2667 MHz DDR4 of memory.
\subsection{BODMAS}%
\subsubsection{Model Performance}
Each of the binary classification models trained on the BODMAS feature dataset achieves high-performance metrics, as indicated by \tabref{tab:train_results_table}. From these metrics, the trained models can accurately distinguish between benign and malicious samples. The high precision value indicates that when a model predicts an instance as malicious (malware), it is likely correct, and the high recall value means that a malware sample is rarely misclassified as benign. While providing valuable immediate feedback on the malware classification abilities, these metrics do not provide insight into the classifier's ability against adversarial examples.  %
%
%
%
%
\begin{table}[tb!]
    \centering
    \caption{Certified robustness accuracy (CRA) and average time to train verification results on 5 samples from each class (125 samples) from the Malimg (family) dataset using NNV and nnenum verification tools on pixel level ($\mathcal{L}_\infty$) perturbations.}
\begin{tabularx}{\columnwidth}{YXXYYY} 
\toprule 
\multirow{2}{*}{Metric} & \multirow{2}{*}{Model} & \multirow{2}{*}{Tool} &\multicolumn{3}{c}{Epsilon ($\epsilon$)}\\
 & & & 1/255 & 2/255 & 3/255 \\
 \midrule 
\multirow{6}{=}{\rotatebox{90}{CRA (\%)}} & \multirow{2}{*}{linear-25} & NNV & 85 & 83 & 79 \\
 &  & nnenum & 90 & 86 & 82 \\
\cline{3-6} 
 \rule{0pt}{3ex} 
 & \multirow{2}{*}{4-25} & NNV & 89 & 76 & 62 \\
 &  & nnenum & 94 & 80 & 66 \\
\cline{3-6} 
 \rule{0pt}{3ex} 
 & \multirow{2}{*}{16-25} & NNV & 88 & 82 & 67 \\
 &  & nnenum & 90 & 86 & 64 \\
\midrule\multirow{6}{=}{\rotatebox{90}{Avg. Time (s)}} & \multirow{2}{*}{linear-25} & NNV & 0.84 & 0.85 & 0.85 \\
 &  & nnenum & 3.60 & 3.63 & 3.69 \\
\cline{3-6} 
 \rule{0pt}{3ex} 
 & \multirow{2}{*}{4-25} & NNV & 17.75 & 41.66 & 82.18 \\
 &  & nnenum & 11.59 & 10.80 & 11.13 \\
\cline{3-6} 
 \rule{0pt}{3ex} 
 & \multirow{2}{*}{16-25} & NNV & 85.00 & 210.00 & 710.25 \\
 &  & nnenum & 38.66 & 44.16 & 43.43 \\
\bottomrule
\end{tabularx}
\end{table}%
\subsubsection{Verification Performance}
\begin{figure}[tb!]
    \centering
    \begin{subfigure}{\columnwidth}
        \centering
        \includegraphics[width=0.72\columnwidth]{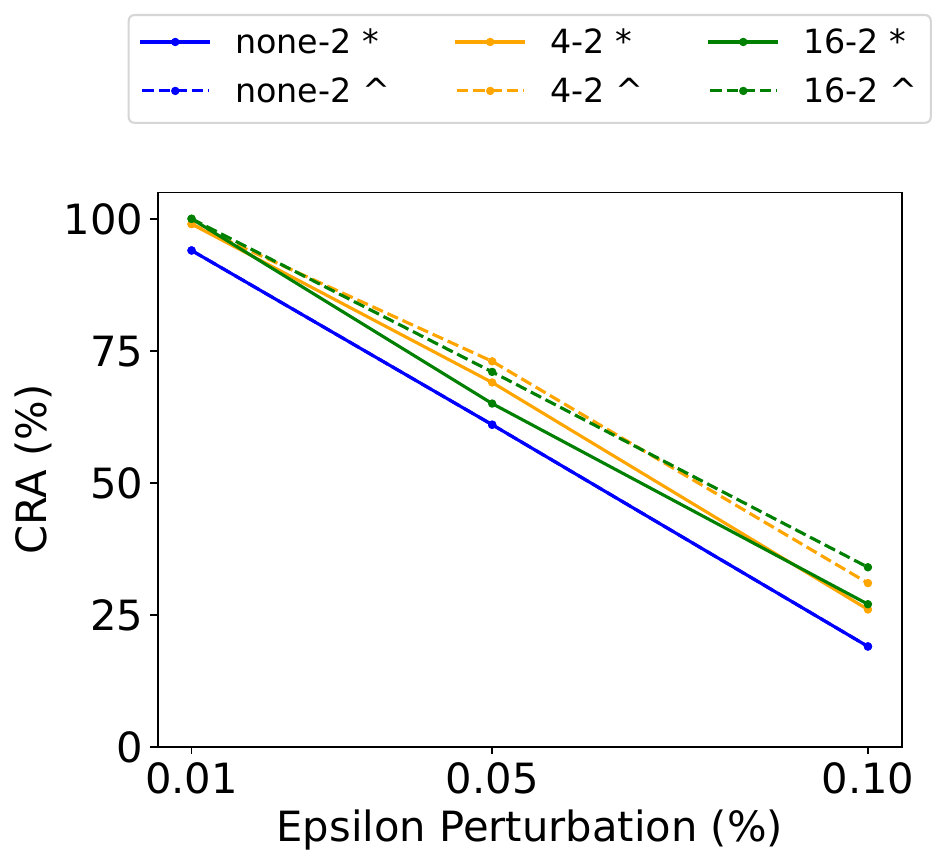}
        \caption{Number Robust}
        \label{fig:bod_res_all}
    \end{subfigure}
    \hfill
    \begin{subfigure}{\columnwidth}
        \centering
        \includegraphics[width=0.72\columnwidth]{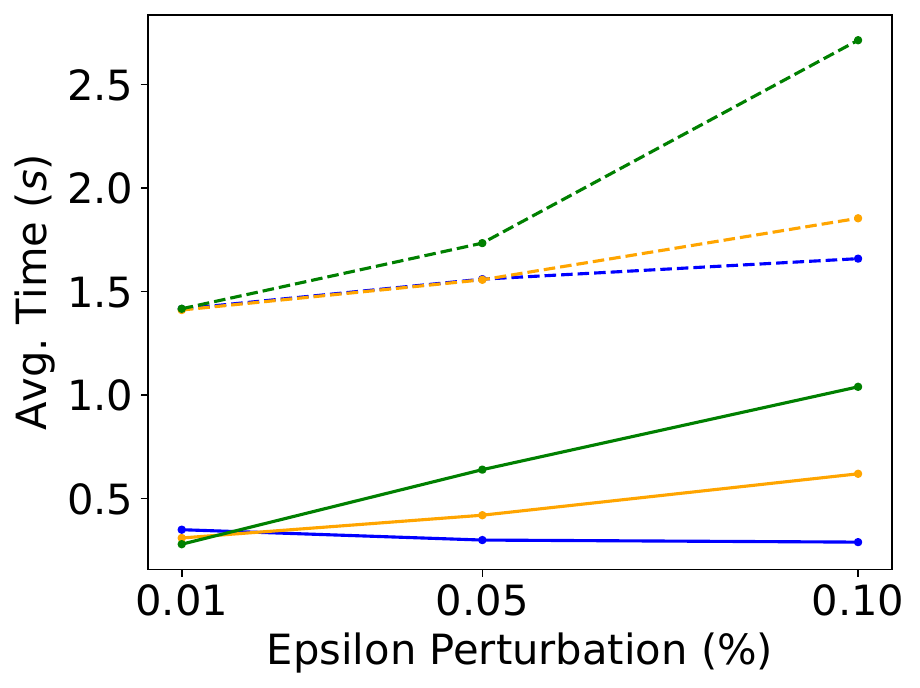}
        \caption{Average Time to Verify}
        \label{fig:bod_time_all}
    \end{subfigure}
    \caption{BODMAS (binary) robustness results ($\mathcal{L}_\infty$) for \textit{all} features using NNV (*) and nnenum ($\wedge$) verification tools. As the perturbation size increases, CRA decreases and time to verify increases. A larger model is more robust to perturbations, especially as the perturbation size increases. While NNV and nnenum have similar robustness accuracy results, NNV takes less time to verify than nnenum for \textit{all} features.}
    \label{fig:bod_all}
\end{figure}

    %
    %
    \textbf{(All)} The certified robustness accuracy (CRA) and average time to verify for the NNV and nnenum verification tools on \textit{all} BODMAS features are shown in \figref{fig:bod_all}. As the perturbation $\epsilon$ increases, the CRA of each of the models (\textit{none-2}, \textit{4-2}, \textit{16-2}) decreases with both NNV and nnenum. Regarding model size, the results indicate that a larger model is more robust to a $\mathcal{L}_\infty$ perturbation, which is increasingly important as the perturbation size increases. While a larger model is more robust, it takes longer to verify, as shown by \figref{fig:bod_time_all}. This makes sense, as a larger perturbation would result in a larger created input set. While not affecting the falsification step, this will affect both the \textit{relax-approx-area} method and \textit{approx-star} method, which is slower for larger models. While the verification tool (NNV vs. nnenum) has little impact on the robustness results, NNV takes less time to verify for each of the perturbation sizes and models.%
   
\begin{figure}[tb!]
    \centering
    \begin{subfigure}{\columnwidth}
        \centering
        \includegraphics[width=0.72\columnwidth]{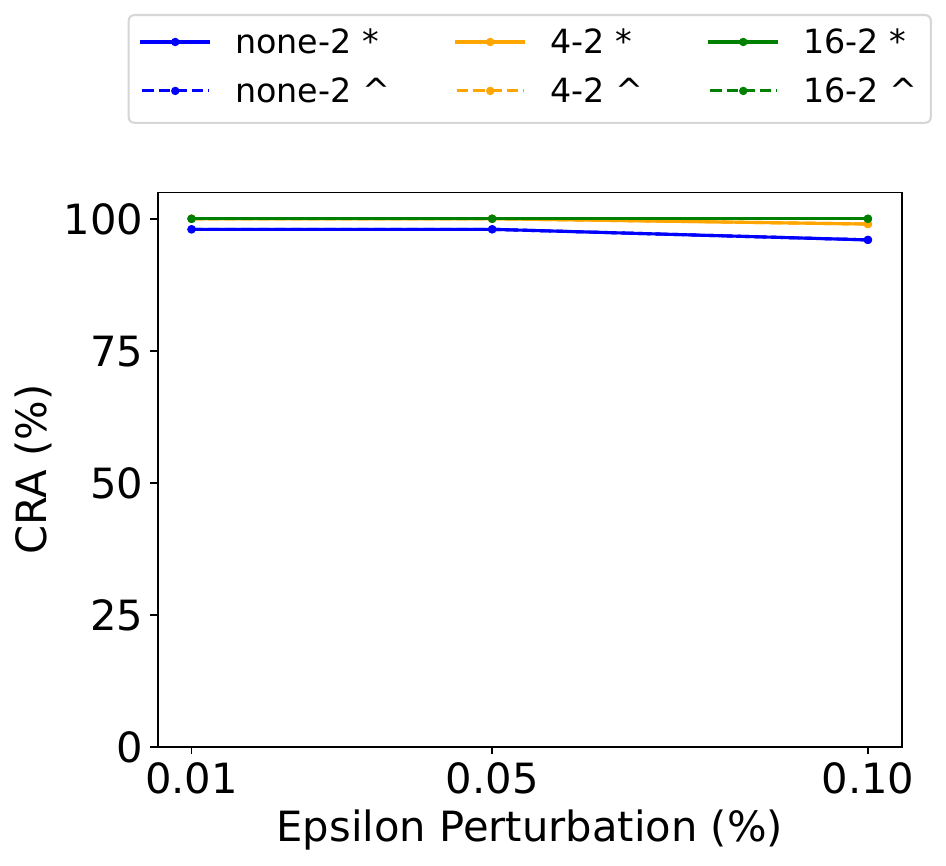}
        \caption{Number Robust}
        \label{fig:bod_res_disc-cont}
    \end{subfigure}
    \hfill
    \begin{subfigure}{\columnwidth}
        \centering
        \includegraphics[width=0.72\columnwidth]{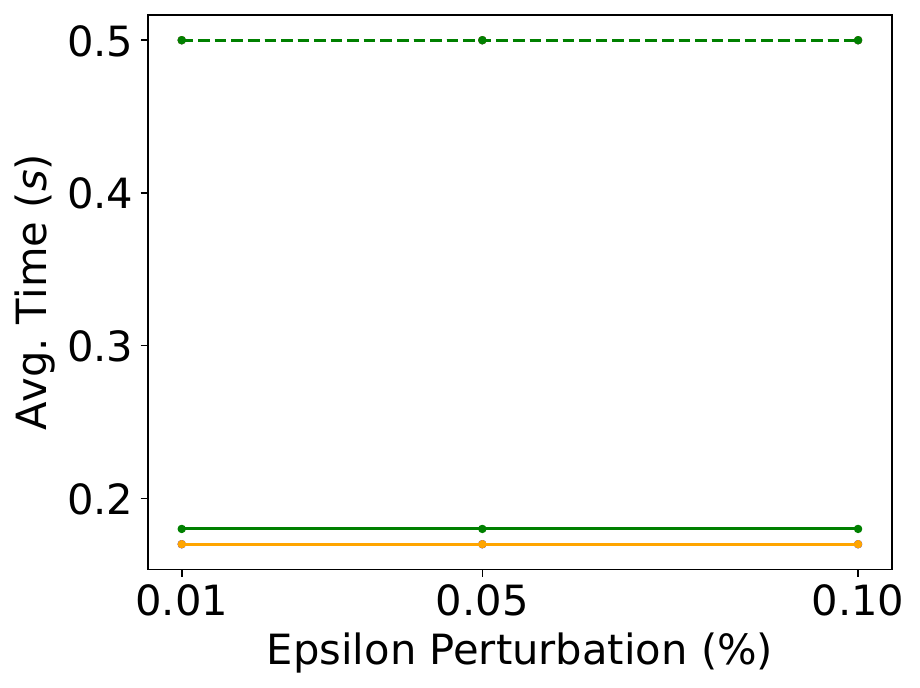}
        \caption{Average Time to Verify}
        \label{fig:bod_time_disc-cont}
    \end{subfigure}
    \caption{BODMAS (binary) robustness results ($\mathcal{L}_\infty$) for \textit{discrete \& continuous} features using NNV (*) and nnenum ($\wedge$) verification tools. Model size does not significantly affect CRA or time to verify for these features. While NNV and nnenum have similar robustness results, NNV takes less time to verify than nnenum for each model and perturbation size.}
    \label{fig:bod_disc-cont}
\end{figure}

    \textbf{(Discrete \& Continuous)} The results of only perturbing \textit{discrete \& continuous} features are shown in \figref{fig:bod_disc-cont}. Whereas model size impacts CRA for perturbations applied to \textit{all} features, when applied to \textit{discrete \& continuous} features, model size does not heavily impact CRA, as seen by the close lines in \figref{fig:bod_res_disc-cont}. Additionally, the time to verify is not significantly impacted by model size either. This experiment highlights the importance of using meaningful features. While the epsilon perturbation sizes for \textit{all} and \textit{discrete \& continuous} are the same (0.01\%, 0.05\%, 0.10\%), the CRA results are drastically different.  For $\epsilon=0.1\%$, 19\% of samples are verified robust for \textit{all} features and 96\% of samples are verified robust for \textit{continuous \& discrete} features. The feature type, therefore, impacts the robustness verification process. Regarding the verification tools, NNV and nnenum reach similar CRA results, but NNV is faster than nnenum for each model and epsilon perturbation size, as demonstrated by the results in \tabref{tab:bodmas_results}. %
        
\begin{figure}[tb!]
    \centering
    \begin{subfigure}{\columnwidth}
        \centering
        \includegraphics[width=0.72\columnwidth]{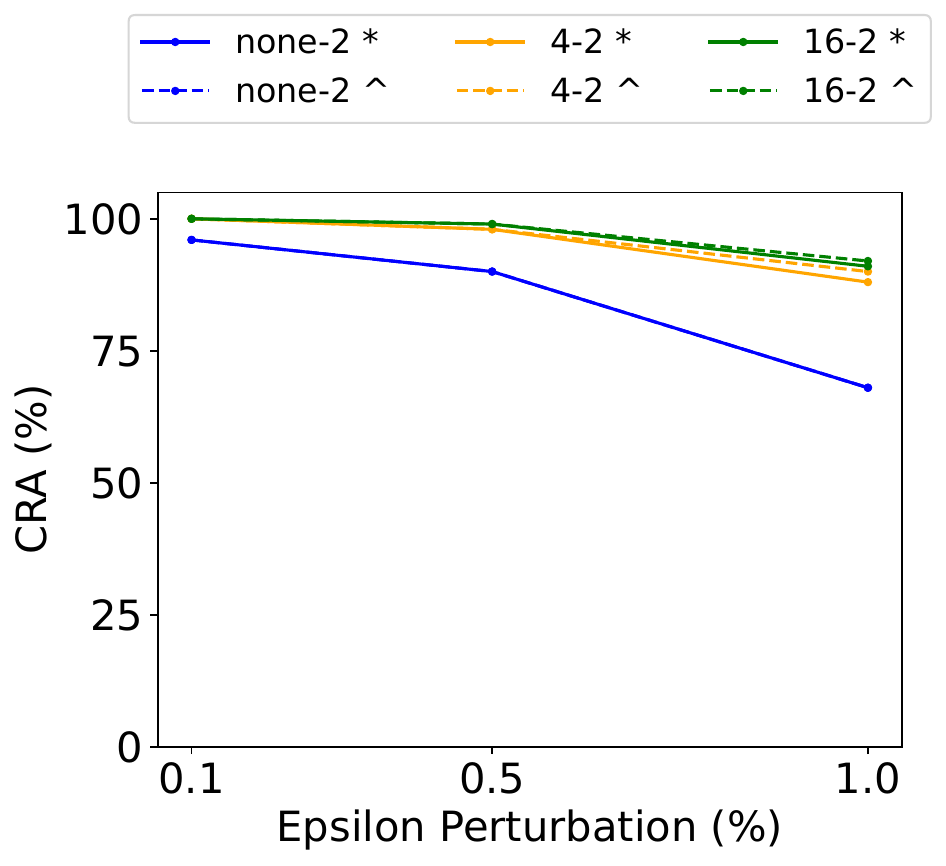}
        \caption{Number Robust}
        \label{fig:bod_res_disc}
    \end{subfigure}
    \hfill
    \begin{subfigure}{\columnwidth}
        \centering
        \includegraphics[width=0.72\columnwidth]{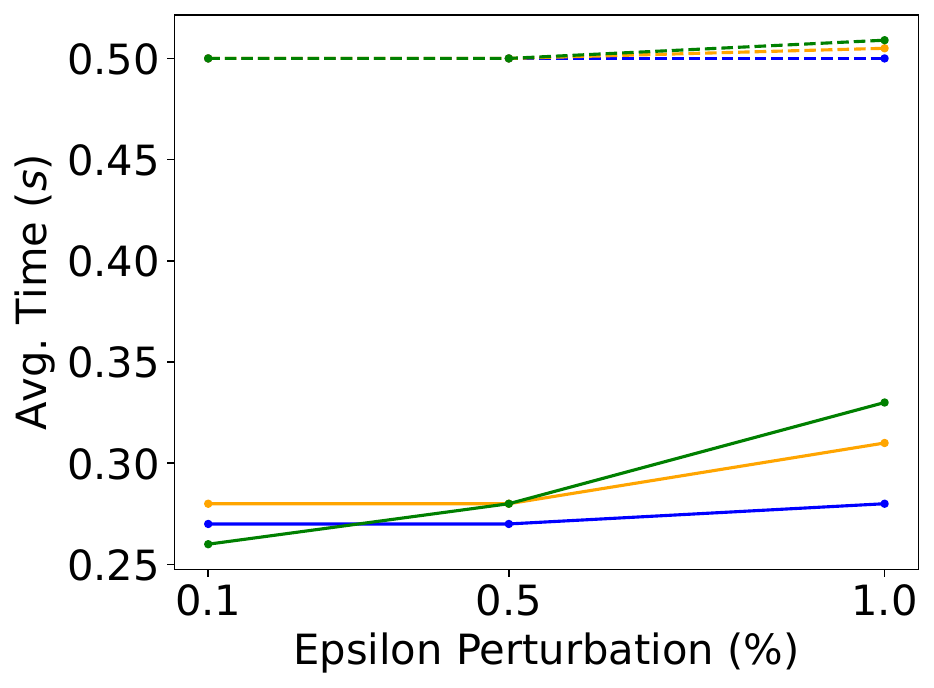}
        \caption{Average Time to Verify}
        \label{fig:bod_time_disc}
    \end{subfigure}
    \caption{BODMAS (binary) robustness results ($\mathcal{L}_\infty$) for \textit{discrete} features using NNV (*) and nnenum ($\wedge$) verification tools. As model size decreases, the model is less robust. While verification results are similar, NNV takes less time to verify than nnenum.}
    \label{fig:bod_disc}
\end{figure}

    \textbf{(Discrete)} The results from applying perturbations to just \textit{discrete} features are shown in \figref{fig:bod_disc}. Whereas model size does not significantly affect CRA for \textit{all} or \textit{discrete \& continuous} features, when applied to \textit{discrete} features only, the smaller the model, the less robust it is. This trend is demonstrated by the low blue line (small model) compared to the higher green and orange lines (larger models). NNV and nnenum garner similar CRA results for \textit{discrete} features, but NNV takes less time to verify, as shown by \figref{fig:bod_time_disc}.

\begin{figure}[tb!]
    \centering
    \begin{subfigure}{\columnwidth}
        \centering
        \includegraphics[width=0.72\columnwidth]{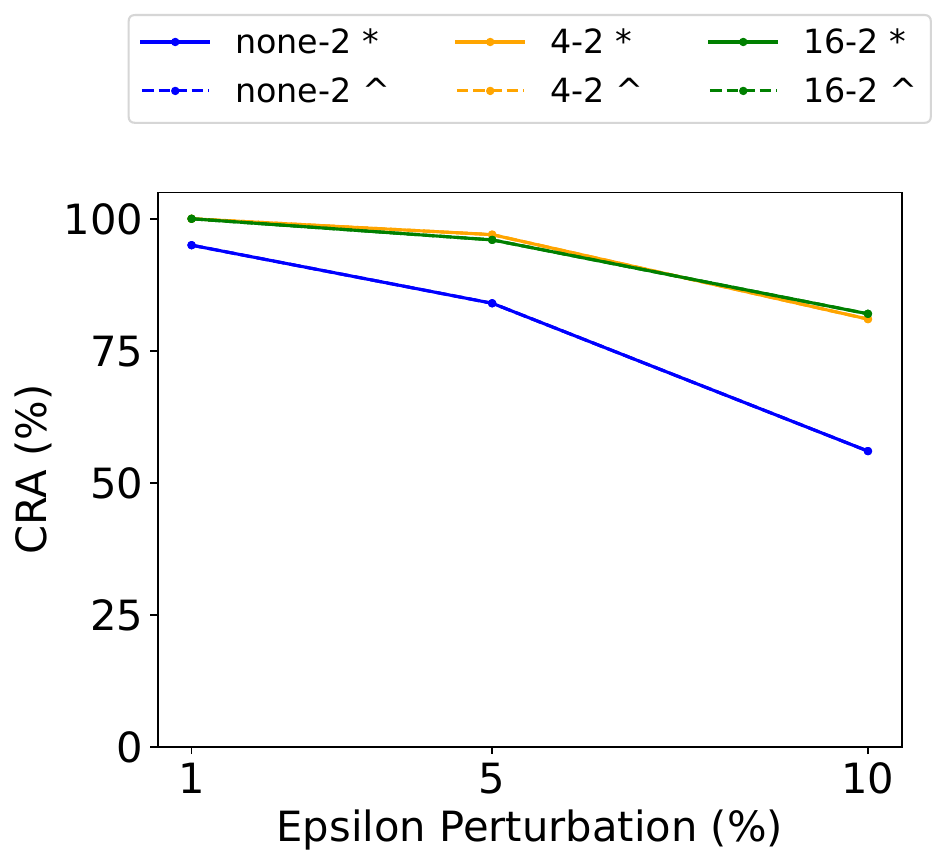}
        \caption{Number Robust}
        \label{fig:bod_res_cont}
    \end{subfigure}
    \hfill
    \begin{subfigure}{\columnwidth}
        \centering
        \includegraphics[width=0.72\columnwidth]{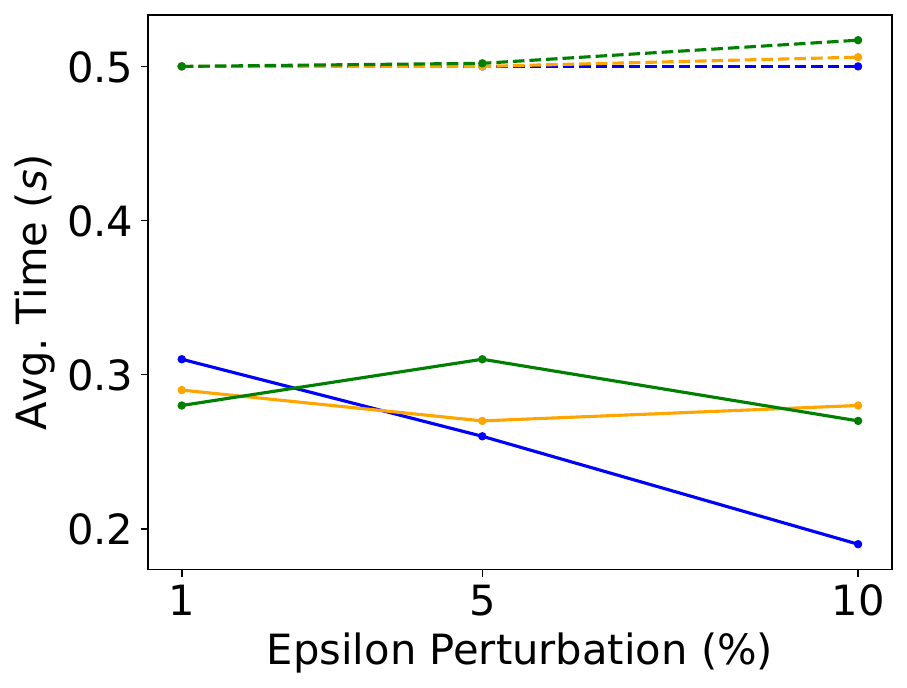}
        \caption{Average Time to Verify}
        \label{fig:bod_time_cont}
    \end{subfigure}
    \caption{BODMAS (binary) robustness results ($\mathcal{L}_\infty$) for \textit{continuous} features using NNV (*) and nnenum ($\wedge$) verification tools. Larger models are more robust as the perturbation size increases. While NNV and nnenum have similar verification results, NNV takes less time to verify.}
    \label{fig:bod_cont}
\end{figure}

    \textbf{(Continuous)} The results of using \textit{continuous} features for verification are shown in \figref{fig:bod_cont}. These results resemble those of \textit{discrete} features, even with the larger perturbation sizes used. This result means that the \textit{discrete} features are more sensitive compared to \textit{continuous} features, as a larger perturbation elicits similar results. When combined (\textit{discrete \& continuous}), however, the CRA remains high (\figref{fig:bod_res_disc-cont}). This is likely due to the smaller epsilon values used for \textit{discrete \& continuous} features. Regarding general trends in \figref{fig:bod_res_disc},  we see that as the perturbation size increases, larger models are more robust. NNV and nnenum also have similar verification results, but NNV takes less time to verify. The average time to verify for the \textit{none-2} model decreases as the perturbation increases, which makes sense when compared to the decreasing CRA. More samples are being falsified early resulting in a shorter verification time. The average verification time for \textit{4-2} and \textit{16-2}, however, remains approximately the same. This is due to an increase in both falsifications (failing early) and an increase in unknown results (failing late), which is keeping the average time about the same even though the CRA is decreasing.
    
\subsection{Malimg}
\subsubsection{Model Performance}
From the training results shown in \tabref{tab:train_results_table}, each of the trained family classifiers is effectively able to discern malware families from known malware samples, as indicated by the high accuracy, precision, recall, and F1 scores. 

\begin{figure}[htb!]
    \centering
    \begin{subfigure}{\columnwidth}
        \centering
        \includegraphics[width=0.72\columnwidth]{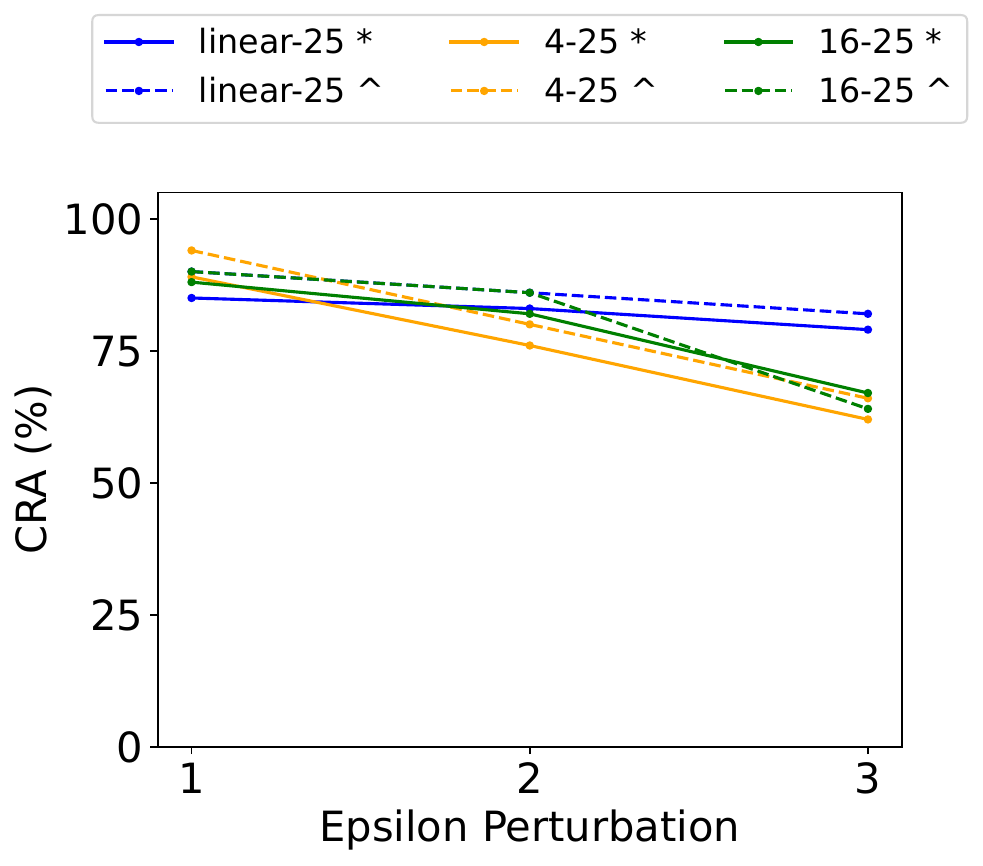}
        \caption{Number Robust}
        \label{fig:malimg_res}
    \end{subfigure}
    \hfill
    \begin{subfigure}{\columnwidth}
        \centering
        \includegraphics[width=0.72\columnwidth]{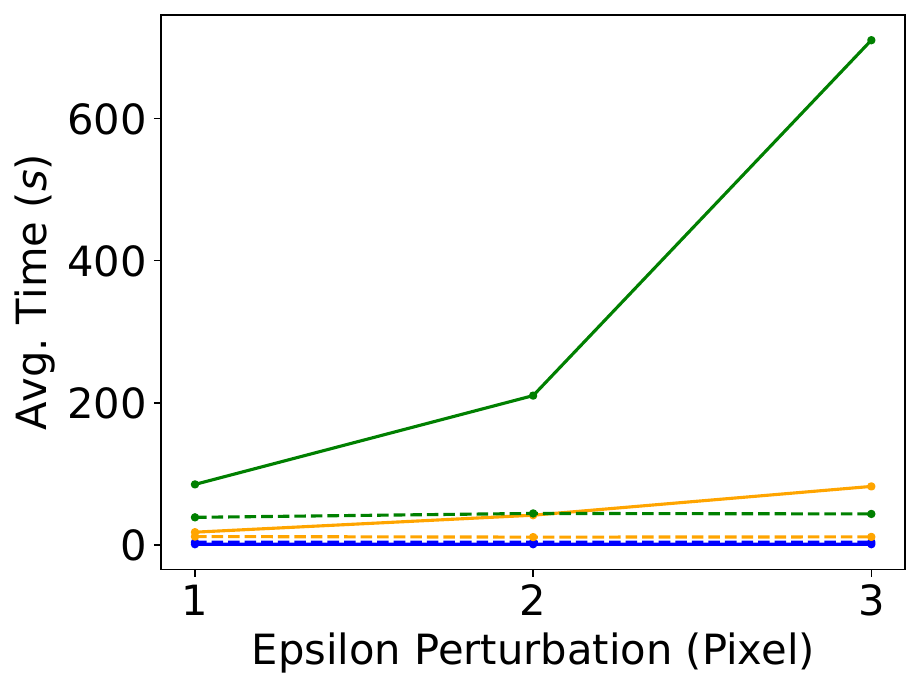}
        \caption{Average Time to Verify}
        \label{fig:malimg_time}
    \end{subfigure}
    \caption{Malimg (family) robustness results ($\mathcal{L}_\infty$) using NNV (*) and nnenum ($\wedge$) verification tools. As perturbation size increases, the CRA decreases for each model. The larger the model, the more time it takes to verify as the perturbation increases.}
    \label{fig:malimg_results}
\end{figure}

\subsubsection{Verification Performance}
The results of the Malimg model verification are shown in \figref{fig:malimg_results}. As the perturbation size increases, the models decrease in CRA, which is especially apparent for the larger models (\textit{4-25} and \textit{16-25}). The larger the model, the more time typically required for each of the verification steps following falsification, as the reachable set for a larger model will be greater than a smaller, linear model. The results in \figref{fig:malimg_results} follow this trend---the larger the model, the more time needed to verify the model, which causes timeouts and a lower CRA value. As a result, smaller models are preferred for this dataset, as they maintain robustness while taking less time to verify. 

\section{Discussion}
As a result of these experiments, we highlight important features for consideration and future study.%

\textbf{\textit{Timing:}} For many of the experiments, NNV resulted in a faster time to verify than nnenum. This is in part due to the different verification methods used by NNV and nnenum. Whereas nnenum is a command line tool requiring only the model and a VNN-LIB file, NNV is a scriptable tool that requires defining the verification steps used during the verification process. Both show the benefit of using each tool: nnenum is a faster process to get started (command line) and NNV is more versatile (adaptable) and, in this case, faster.


\textbf{\textit{Importance of Features:}} In the BODMAS dataset, making feature specific perturbations makes a significant difference in the robustness evaluations of the model, as shown by the difference between \figref{fig:bod_res_all} and \figref{fig:bod_res_disc-cont}. More work is therefore necessary to determine meaningful features and meaningful perturbations to those features. 

\textbf{\textit{Sample Coverage:}} The verification sets for both Malimg and BODMAS contain inputs that rarely, if ever, generate a `robust' outcome. \tabref{tab:malimg_per_class} shows the per class results of verification using NNV on the samples perturbed with $\epsilon=2$. \textit{Autorun.K}, \textit{Swizzor.gen!}, and \textit{Swizzor.gen!I} have a low number of robust samples for each trained model. While these classes are underrepresented during training compared to classes like \textit{Allaple.A}, other underrepresented classes do not have similar results. For instance, \textit{Skintrim.N} only has 55 training samples, but images from this class are robustly verified 5/5 times for each model. These results highlight that some classes are more difficult to verify than others. It might be better to evaluate malware classifier models on levels of inputs rather than a conglomerate. For instance, model X is robust to levels 1, 2, and 3 but not to level 4. This might provide more valuable insight into the robustness of a model. This level-specific result will be an interesting problem to explore in future works. %
\begin{table}[]
    \centering
\begin{tabular}{ccccc}
\toprule 
\multirow{2}{*}{Malimg Class} & \multicolumn{3}{c}{No. Certified Robust} & \multirow{2}{*}{Train Samples}\\
& linear-25 & 4-25 & 16-25 & \\
\midrule 
Adialer.C & 5 & 5 & 5 & 97\\
Agent.FYI & 5 & 5 & 5 & 91\\
Allaple.A & 5 & 5 & 5 & 2824\\
Allaple.L & 5 & 4 & 4 & 1491\\
Alueron.gen!J & 5 & 5 & 5 & 173\\
Autorun.K & 0 & 0 & 0 & 81\\
C2LOP.P & 4 & 2 & 1 & 121\\
C2LOP.gen!g & 5 & 3 & 4 & 175\\
Dialplatform.B & 5 & 5 & 5 & 152\\
Dontovo.A & 5 & 5 & 5 & 137\\
Fakerean & 5 & 5 & 5 & 306\\
Instantaccess & 5 & 5 & 5 & 356\\
Lolyda.AA1 & 3 & 5 & 5 & 153\\
Lolyda.AA2 & 5 & 5 & 5 & 159\\
Lolyda.AA3 & 3 & 5 & 5 & 98\\
Lolyda.AT & 5 & 2 & 5 & 134\\
Malex.gen!J & 3 & 0 & 3 & 111\\
Obfuscator.AD & 5 & 5 & 5 & 117\\
Rbot!gen & 5 & 5 & 5 & 133\\
Skintrim.N & 5 & 5 & 5 & 55\\
Swizzor.gen!E & 2 & 0 & 0 & 103\\
Swizzor.gen!I & 0 & 0 & 0 & 107\\
VB.AT & 5 & 5 & 5 & 383\\
Wintrim.BX & 4 & 4 & 5 & 72\\
Yuner.A & 5 & 5 & 5 & 775\\
\bottomrule 
 \end{tabular}
    \caption{Per class robustness accuracy on the Malimg verification set for the $\epsilon=2$ perturbation using NNV. Some classes are more difficult to verify than others, as highlighted by the \textit{Autorun.K}, \textit{Swizzor.gen!E},  and \textit{Swizzor.gen!I} classes.}
    \label{tab:malimg_per_class}
\end{table}%
\section{Related Work}
\textbf{\textit{Malware Classification:}} Previous works have shown the promise of using machine learning models for malware classification. In \cite{schultz2000data,kolter2004learning}, the authors use data mining to extract meaningful features, and then apply traditional machine learning algorithms. Recent research has shifted to the use of neural networks. In \cite{nataraj2011malware}, the authors introduce the use of image malware classification, i.e., converting binaries into images and utilizing convolution neural networks (CNNs) for binary and/or family classification. This approach has since been refined by numerous works, including \cite{vasan2020image}.  

\textbf{\textit{Semantically Meaningful Perturbations:}} Semantically meaningful perturbations have been addressed in recent work related to images. In \cite{paterson2021deepcert}, the authors investigate contextually meaningful perturbations to deep neural network image classifiers on `German Traffic Sign' and `CIFAR-10' data. In \cite{uzunova2019interpretable}, the authors utilize a domain `deletion` perturbation to medical image pathologies.

\textbf{\textit{Neural Network Verification:}} The area of DNN verification has increasingly grown in recent years, leading to the development of standard input formats\footnote{vnnlib: \url{https://www.vnnlib.org}} \cite{vnnlib23} as well as friendly competitions~\cite{manzanas2022archcomp,muller2022vnncomp}, that help compare and evaluate all the recent methods and tools proposed in the community~\cite{wang2021beta,wang2018reluval,katz2017reluplex,katz2019marabou,bak2021nfm}. The majority of these methods focus on verifying FFNN and CNN architectures.
These approaches can generally be classified into sound or unsound, and complete or incomplete. 
Unsound refers to probabilistic analysis such as~\cite{tran2023probstar}, although they are less common for NN verification than sound approaches. 
Complete and sound methods refer to algorithms that can precisely analyze whether a given property holds on a model, also referred to as exact methods. However, these type of methods suffer from scalability issues as the exact analysis tend to be very computationally expensive. These are also limited in the type of layers and model architectures they can be applied to. These can be Satisfiability Modulo Theories (SMT) based methods~\cite{katz2019marabou,katz2017reluplex}, Mixed Integer Linear Program (MILP) based methods~\cite{tjeng2017mipverify}, reachability analysis methods~\cite{tran2021cav}, and others such as branch and bound methods~\cite{bunel2020bab}.
Incomplete methods refer to sound approaches that present a tradeoff between precision and scalability, which allows a faster computation of a verification problem of larger NNs than sound and complete methods, although it can lead to \emph{unknown} results due to the approximation used. Several of these methods are based on abstract interpretation, some of which have demonstrated to outperform complete methods by orders of magnitude (time wise)~\cite{muller2022vnncomp}. Recent work in \cite{elboher2022neural} has enhanced the abstraction-based verification of neural networks via residual reasoning.

\section{Conclusion}
In this paper, we present a case study to introduce and evaluate the novel formal verification of malware classification for family identification as well as malware identification. Through testing, training, and verification processes, we highlight important current malware verification capabilities as well as areas to be considered moving forward. Through this rigorous analysis, we hope to increase the visibility of this verification domain, bolstering refinement to a safety-critical system that has a drastic impact on our lives every day. 

\begin{acks}
This paper was supported in part by a fellowship award under contract FA9550-21-F-0003 through the National Defense Science and Engineering Graduate (NDSEG) Fellowship Program, sponsored by the Air Force Research Laboratory (AFRL), the Office of Naval Research (ONR), and the Army Research Office (ARO). The material presented in this paper is based upon work supported by the National Science Foundation (NSF) through grant numbers 2220426 and 2220401, the Defense Advanced Research Projects Agency (DARPA) under contract number FA8750-23-C-0518, and the Air Force Office of Scientific Research (AFOSR) under contract number FA9550-22-1-0019 and FA9550-23-1-0135. Any opinions, findings, and conclusions or recommendations expressed in this paper are those of the authors and do not necessarily reflect the views of AFOSR, DARPA, or NSF.
\end{acks}

\bibliographystyle{ACM-Reference-Format}
\bibliography{references, diego, kleach}

\end{document}